\documentclass[aps,epsf,prb]{revtex4}
\usepackage{graphicx}
 \usepackage{epsfig}
\newcommand{\av}[1]{ {\left\langle #1 \right\rangle} }

\newcommand{\avpo}[1]{ {\left\langle #1 \right\rangle}_{N_1+1} }
\newcommand{\avpt}[1]{ {\left\langle #1 \right\rangle}_{N_2+1} }

\newcommand{\tpo}{\langle\tilde{P}\rangle_{N_1+1}}
\newcommand{\tpt}{\langle \tilde{P} \rangle_{N_2+1}}

\newcommand*{\nd}{\text{ and}}

\begin{document}

\title[Two SET - Resonator Dynamics and Noise]{Noise properties of two single electron transistors coupled by a nanomechanical resonator}%
\author{D.A. Rodrigues and A.D. Armour}%
\affiliation{School of Physics and Astronomy, University of Nottingham,
Nottingham NG7 2RD, United Kingdom}%

\begin{abstract}
We analyze the noise properties of two single electron transistors
(SETs) coupled via a shared voltage gate consisting of a
nanomechanical resonator.  Working in the regime where the
resonator can be treated as a classical system, we find that the
SETs act on the resonator like two independent heat baths. The
coupling to the resonator generates positive correlations in the
currents flowing through each of the SETs as well as between the
two currents. In the regime where the dynamics of the resonator is
dominated by the back-action of the SETs,  these positive
correlations can lead to parametrically large enhancements of the
low frequency current noise. These noise properties can be
understood in terms of the effects on the SET currents of
fluctuations in the state of a resonator in thermal equilibrium
which persist for times of order the resonator damping time.
\end{abstract}
\maketitle

\section{Introduction}

Nanoelectromechanical systems in which micron-sized mechanical
resonators couple to the transport electrons of a nearby conductor
form a new and interesting class of mesoscopic system.\cite{nems}
In particular, there has been considerable
theoretical\cite{white,mmh,ABZ,ADA,bruder,Clerk1,Clerk2,blencowe}
and experimental\cite{mset1,mset2} interest in the properties of
nanomechanical single electron transistors in which a mechanical
resonator forms the voltage gate of the transistor. For these
systems, the conductance properties of the SET are extremely
sensitive to the resonator motion and such a device has been used
to measure the displacement of a nanomechanical resonator with
almost quantum limited precision.\cite{mset2}

Apart from its application as an ultra-sensitive displacement
detector, the SET-resonator system has a number of interesting
features arising from its coupled dynamics. As electrons pass
through the island of the SET they exert a stochastic driving
force on the resonator, but the motion of the resonator in turn
affects the rates at which electrons tunnel between the leads and
the island of the SET, leading to non-trivial correlations between
the electrical and mechanical motion. In general the resonator-SET
system has a complicated coupled dynamics, but in the
experimentally relevant regime of relatively large applied bias,
but low electromechanical coupling and resonator frequency, the
effect of the electrons on the mechanical resonator turns out to
be closely analogous to that of an equilibrium thermal
bath.\cite{ABZ,blencowe}

The correlations between the electrical and mechanical degrees of
freedom in nanomechanical SETs also give rise to a number of
unusual features in the current noise spectrum of the SET. There
is a strong enhancement of the current noise at the resonator
frequency, there can also be a strong enhancement at the first
harmonic of the resonator frequency and at low
frequencies.\cite{ADA} Similar features have also been predicted
in the noise spectra of a number of closely related
nanoelectromechanical
systems.\cite{isac,bruder,novotny,Clerk1,Clerk2} Of particular
interest in such systems is the unusual behavior of the
zero-frequency current noise, which can become parametrically
large when the resonator is under damped.\cite{isac,novotny}

Enhancement of the zero-frequency current noise is also known to
occur, under certain circumstances, in systems consisting of two
parallel SETs, or quantum dots, with a direct electrostatic
interaction between the two islands. Such systems have been
investigated quite extensively,\cite{set1,set2,set3,qd1} and it
has been shown that the electrostatic interactions between charges
on the two island give rise to important cross-correlations
between the currents in the individual SETs\cite{set1} which can
generate either positive or negative correlations between the
carriers and hence either enhance or suppress the noise, depending
on the exact details of the system.\cite{set1,set2,set3,qd1} Apart
from their intrinsic interest, correlations between the currents
of two SETs can be used to enhance the sensitivity of charge
detection.\cite{buehler}

In this paper we investigate the noise properties of a system
consisting of a nanomechanical resonator coupled to two SETs
aligned in parallel and relate them to the dynamics of the
resonator. The resonator is assumed to lie between the islands of
the two SETs acting as a mechanically compliant voltage gate for
both of them. In order to act as a gate, the resonator is coated
with a thin metal layer and kept at a fixed voltage. Under such
circumstances there is no direct electrostatic interaction between
the SET islands, but the mutual interaction between the electrons
travelling through the SETs and the resonator nevertheless
generates correlations between the currents flowing in the two
conductors.

As electrons pass through the SETs they exert a stochastic force
on the resonator which can strongly affect its motion, but the
motion of the resonator in turn affects the motion of electrons
through the SETs. We find that the dynamical state of a resonator
coupled to two SETs is approximately equivalent to that of an
oscillator coupled to two independent thermal baths. We also find
that the presence of the resonator strongly enhances the low
frequency current noise of the individual SETs and generates
positive correlations between the currents in the two SETs. The
correlations in the currents within each of the individual SETs,
and between the currents in the two SETs, are generated by
fluctuations in the state of the resonator and hence the magnitude
of the low frequency noise depends sensitively on the time-scale
over which they decay, the damping time of the resonator. Although
the fluctuations in the state of the resonator are in turn
generated by the motion of the electrons through the SETs, the
resulting noise properties are very similar to those obtained if
the back-action of the SETs on the resonator is neglected and
instead the resonator is assumed to be in a fixed thermal state
with appropriately chosen parameters.

The outline of this paper is as follows.  In Section \ref{sec:ms}
we describe our model for the two-SET resonator system and the
conditions under which it is valid. We also introduce the master
equation formalism which is used to derive the subsequent results.
Then in Section \ref{sec:meq} we analyze the dynamics of the
resonator when coupled to two SETs and compare the results to
those obtained for a resonator coupled to a single SET.  In
Section \ref{sec:current} we investigate the noise properties of
the SET currents. First, we present calculations of the
zero-frequency current noise in one of the SETs and relate the
results to a simple model where the back-action of the SETs on the
resonator is neglected. Then we describe the cross-correlations
between the currents in the two SETs, and how they can be
investigated via measurements on the combined currents through the
two leads. Finally, in Section \ref{sec:conclusion} we present our
conclusions. The Appendix contains additional details about how an
effective equation for the resonator is derived.

\section{Model System} \label{sec:ms}

\begin{figure}
\noindent
\epsfig{figure=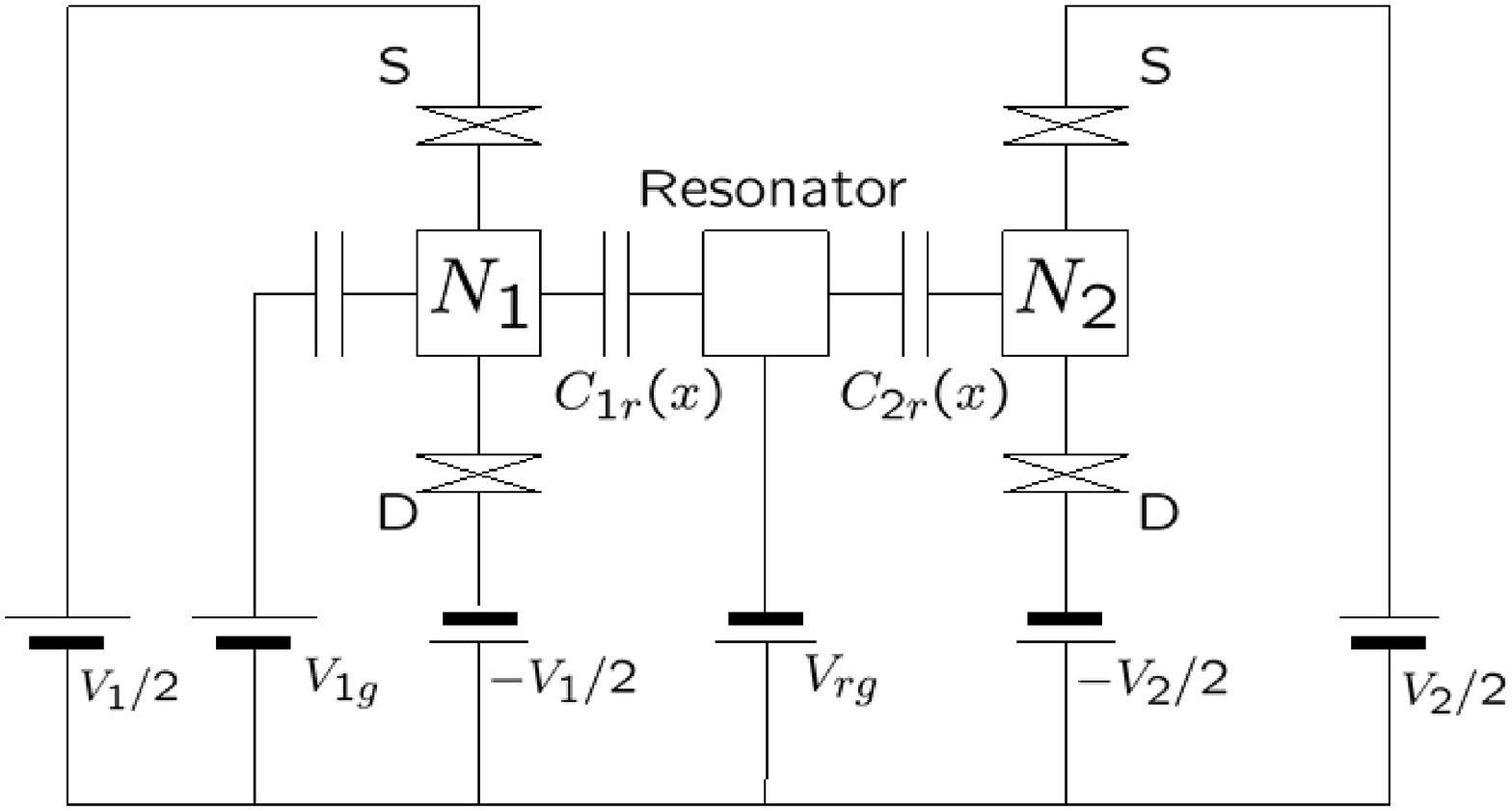,width=10cm} \caption{Schematic Circuit
diagram for the 2 SET resonator system. The central box represents
the resonator and the outer boxes  represent the two SET islands
containing $N_1$ and $N_2$ electrons.} \label{fig:2SETdiag}
\end{figure}

The system of two SETs coupled to a nanomechanical resonator which
we consider is illustrated schematically in Fig.\
\ref{fig:2SETdiag}. The SET islands, labelled 1 and 2, are linked
to voltage leads by source (S) and drain (D) junctions across
which voltages $V_{1}$ and $V_{2}$ respectively are applied. The
resonator acts as a movable gate for island 1(2) with capacitance
$C_{1(2)r}(x)$ which depends on its position, $x$. In practice the
motion of the resonator will be very small on the scale of the
equilibrium separation between it and the SET islands, $d_{1(2)}$,
and hence can be treated as a linear perturbation,
$C_{1(2)r}(x)=C_{1(2)r}(1+x/d_{1(2)})$. We also assume that SET 1
has an additional gate with a capacitance $C_{1g}$ and voltage
$V_{1g}$ so that the operating points of the two SETs can be tuned
independently.

The dynamics of the system can be described using a generalization
of the classical master equation formalism which was used to
analyze the coupled dynamics of a nanomechanical resonator coupled
to one SET.\cite{ABZ,ADA} The basic assumptions underlying this
approach are that the charge state of the SET islands are each
limited to two possible values by charging effects, with
transitions between these charge states occurring via electron
tunnelling processes which are adequately described by the
orthodox model,\cite{ferry} and that the resonator can be treated
as a classical harmonic oscillator. These conditions are met if
the background temperature is much lower than the charging energy
of the SET islands, the SETs are assumed to tuned to, or close to,
degeneracy and the drain-source voltages applied are much larger
than the quantum of energy associated with the
resonator.\cite{ABZ}

Since the electronic motion is stochastic, the system may be
described by a set of probability distributions of the form
$P^{n_1,n_2}_{n,m}(x,u;t)$, which give the probability at time $t$
of having $n$ excess electrons on the island of SET 1, $m$ excess
electrons on the island of SET 2 and the resonator being at a
position $x$ with velocity $u$. The superscripts on the
distributions, $n_{1(2)}$, are count variables which give the
number of electrons that have passed through the source junction
of SET 1(2). Although the count variables have no effect on the
resonator dynamics, they play an important role in the analysis of
the noise properties of the SETs.

The resonator is assumed to be a simple harmonic oscillator with
frequency $\omega_0$ and effective mass $m$. The quantum of energy
associated with the resonator, $\hbar \omega_0$, is assumed to be
the smallest energy in the problem, so that it is always much less
than the energies associated with the voltages applied across the
SET islands, $eV_1$ and $eV_2$, to ensure that quantum effects in
the resonator's dynamics can be neglected. The relevant charge
states for the SETs are assumed to have $N_{i}$ and $N_{i}+1$
excess electrons on island $i=1,2$. We choose the origin of
resonator position $x$ such that $x=0$ is the equilibrium position
of the resonator (i.e.\ the minimum in its potential) when there
are $N_1$ and $N_2$ electrons on SET islands 1 and 2,
respectively. If there are $N_{1(2)}+1$ electrons on island 1(2),
the equilibrium position of the resonator is shifted by the
electrostatic force to $x=x_{1(2)}$, and if both islands contain
an extra electron, then the equilibrium position of the resonator
is shifted by $x=x_1+x_2$. The lengths $x_1$ and $x_2$ are
essentially measures of the strength of the interaction between
the SETs and the resonator, they are given by
$x_{1(2)}=-\frac{|e|C_{1(2)r}V_{rg}}{(d_{1(2)}C_{1(2)\Sigma}m
\omega_0^2)}$ where $C_{1(2)\Sigma}$ is the total capacitance of
SET island 1(2) and $d_{1(2)}$ is the separation between between
SET island 1(2) and the resonator measured along the direction of
increasing $x$ (hence the signs of $x_1$ and $x_2$ will be
opposite for the setup in Fig. \ref{fig:2SETdiag}).

Transitions between the charge states of the SET islands are
described by the rates for electron tunnelling forwards (positive
charge moving from source to drain), $+$, or backwards, $-$,
through the source (drain) tunnel junctions of islands $1$ or $2$,
which are written as  $\Gamma_{1S(1D)}^\pm$ and
$\Gamma_{2S(2D)}^\pm$. According to the orthodox
model,\cite{ferry} these electron tunnelling rates depend on the
differences in electrostatic energy of the system before and after
the tunnelling events. The resonator affects the tunnel rates
because its motion changes the gate capacitance of the SET islands
and hence the electrostatic energy differences. Assuming that the
thermal energies of the electrons in the SETs, and the energies
associated with the electromechanical couplings,
$m\omega_0^2x_{1(2)}^2$, are both much less than the energy scale
set by the voltages, $eV_{1(2)}$, then the backward tunnel rates
can be set to zero and the forward tunnel rates can be written in
an intuitive way as\cite{ABZ,ADA}
\begin{eqnarray}
\Gamma^+_{1(2)S}&=&\Gamma_{1(2)S}-\frac{m\omega_0^2x_{1(2)}x}{e^2R_{1(2)S}}\\
\Gamma^+_{1(2)D}&=&\Gamma_{1(2)D}+\frac{m\omega_0^2x_{1(2)}x}{e^2R_{1(2)D}}
\end{eqnarray}
where $\Gamma_{1(2)S},\Gamma_{1(2)D}$ are the position-independent
parts of the tunnel rates, and $R_{1(2)S(D)}$ is the resistance of
the relevant SET tunnel junction. Notice that we have assumed that
the coupling is weak enough that it can be treated
linearly.\cite{ABZ}

We are now in a position to set up the master equations for the
system. However, the overall number of parameters describing the
system remains rather large, making it difficult to extract the
essential features of the dynamics. Hence, we will make a few
further simplifying assumptions. We choose to consider only the
case where the couplings of the SETs to the resonator are equal
and opposite, $x_2=-x_1$, and we assume that the resistances of
all the tunnel junctions have the same value, $R$. Finally, we
assume that the SETs are tuned to the current peaks through the
gate voltages so that the position independent parts of the
tunnelling rates through the source and drain junctions of each
SET are equal. As long as the charging energies are larger than
the voltages ($e^2/2C_{1(2)\Sigma}\gtrsim eV_{1(2)}$), the
orthodox model predicts tunnel rates at the current peak which are
proportional to the source-drain
voltages:\cite{korotkov_94,korotkov_96}
$\Gamma_{1(2)S}=\Gamma_{1(2)D}=V_{1(2)}/(2eR)$.

The most obvious way to describe how the state of the system
evolves would be to derive master equations for the four
distributions $P^{n_1,n_2}_{N_1,N_2}(x,u;t)$,
$P^{n_1,n_2}_{N_1+1,N_2}(x,u;t)$, $P^{n_1,n_2}_{N_1,N_2+1}(x,u;t)$
and $P^{n_1,n_2}_{N_1+1,N_2+1}(x,u;t)$. However, in practice it
proves more convenient to work with an alternative set of
distributions composed of
\begin{eqnarray}
P^{n_1,n_2}(x,u;t) &=&P^{n_1,n_2}_{N_1,N_2}(x,u;t)
+P^{n_1,n_2}_{N_1+1,N_2}(x,u;t)+P^{n_1,n_2}_{N_1,N_2+1}(x,u;t)
+P^{n_1,n_2}_{N_1+1,N_2+1}(x,u;t)\\
P^{n_1,n_2}_{N_1+1}(x,u;t)
&=&P^{n_1,n_2}_{N_1+1,N_2}(x,u;t)+P^{n_1,n_2}_{N_1+1,N_2+1}(x,u;t)\\
P^{n_1,n_2}_{N_2+1}(x,u;t)
&=&P^{n_1,n_2}_{N_1,N_2+1}(x,u;t)+P^{n_1,n_2}_{N_1+1,N_2+1}(x,u;t),
\end{eqnarray}
and $P^{n_1,n_2}_{N_1+1,N_2+1}(x,u;t)$. These distributions are
found to evolve according to the following set of master
equations, written in dimensionless form:
\begin{eqnarray}
\frac{\partial P^{n_1,n_2}}{\partial t} &=& \epsilon^2 x
\frac{\partial P^{n_1,n_2}}{\partial u}-u \frac{\partial
P^{n_1,n_2}}{\partial x}-\epsilon^2  \frac{\partial}{\partial
u}\left[ P^{n_1,n_2}_{N_1+1}- P^{n_1,n_2}_{N_2+1}\right]
+\left(\frac{\Gamma_{1}}{2}-\kappa \,
x\right)\left[P^{n_1-1,n_2}_{N_1+1}-P^{n_1,n_2}_{N_1+1}\right]\nonumber\\
&&+\left(\frac{\Gamma_{2}}{2}+\kappa \, x
\right)\left[P^{n_1,n_2-1}_{N_2+1}-P^{n_1,n_2}_{N_2+1}\right]\label{eq:motionP}\\
 \frac{\partial P^{n_1,n_2}_{N_1+1}}{\partial t} &=&  \epsilon^2(x-1)
\frac{\partial P^{n_1,n_2}_{N_1+1}}{\partial u}+\epsilon^2
\frac{\partial P^{n_1,n_2}_{N_1+1,N_2+1}}{\partial u}-u
\frac{\partial P^{n_1,n_2}_{N_1+1}}{\partial x}+ \left(
\frac{\Gamma_{1}}{2}+ \kappa\, x \right)P^{n_1,n_2}
-\Gamma_{1} \, P^{n_1,n_2}_{N_1+1}\nonumber\\
&&+\left(\frac{\Gamma_{2}}{2} +\kappa\, x
\right)\left[P^{n_1,n_2-1}_{N_1+1,N_2+1}-P^{n_1,n_2}_{N_1+1,N_2+1}\right]
 \label{eq:PN1} \\
\frac{\partial
 P^{n_1,n_2}_{N_2+1}}{\partial
t} &=&  \epsilon^2(x+1) \frac{\partial
P^{n_1,n_2}_{N_2+1}}{\partial u}-\epsilon^2\frac{\partial
P^{n_1,n_2}_{N_1+1,N_2+1}}{\partial u}-u \frac{\partial
P^{n_1,n_2}_{N_2+1}}{\partial x}+  \left( \frac{\Gamma_{2}}{2}-
\kappa\, x\right)P^{n_1,n_2} -\Gamma_{2}P^{n_1,n_2}_{N_2+1}
\nonumber\\
&&+\left(\frac{\Gamma_{1}}{2} -
\kappa\, x\right)\left[P^{n_1-1,n_2}_{N_1+1,N_2+1}-P^{n_1,n_2}_{N_1+1,N_2+1}\right] \label{eq:PN2} \\
\frac{\partial
 P^{n_1,n_2}_{N_1+1,N_2+1}}{\partial
t} &=& \epsilon^2 x \frac{\partial P^{n_1,n_2}_{N_1+1,N_2+1}
}{\partial u} -u \frac{\partial P^{n_1,n_2}_{N_1+1,N_2+1}
}{\partial x}+ \left( \frac{\Gamma_{1}}{2}+ \kappa\,
x\right)P^{n_1,n_2}_{N_2+1}+ \left( \frac{\Gamma_{2}}{2}-
\kappa\, x\right)P^{n_1,n_2}_{N_1+1}\nonumber\\
&&-(\Gamma_{1}+\Gamma_2)\,P^{n_1,n_2}_{N_1+1,N_2+1}
\label{eq:PN12},
\end{eqnarray}
where position and time have been scaled by $x_1$ and
$eR/V=1/\Gamma$ respectively, with the average voltage
$V=(V_1+V_2)/2$ introduced to preserve the natural symmetry of the
equations. The dimensionless electromechanical coupling is given
by $\kappa={m\omega_0^2x_{1}^2}/(eV)$, the resonator frequency is
expressed in dimensionless units as $\epsilon=\omega_0/\Gamma$,
and the dimensionless tunnel rates, $\Gamma_1,\Gamma_2$, are
defined by the relations: $\Gamma_1/2=\Gamma_{1S(D)}/\Gamma$ and
$\Gamma_2/2=\Gamma_{2S(D)}/\Gamma$.

The coupled set of master equations contain all the information
about the dynamics of the system and can be used to extract
information both about the motion of the resonator and about the
noise properties of the SETs. In exploring the behavior of the
system we will concentrate on the effects of varying three
parameters: the reduced frequency of the resonator, $\epsilon$,
the electromechanical coupling $\kappa$ and the ratio of SET
voltages $V_1/V_2$ (which also changes $\Gamma_1/\Gamma_2$).
However, because of the underlying assumptions in the model these
parameters cannot be varied arbitrarily. In particular, the range
of values for the voltages $V_1$ and $V_2$ is bounded below, and
the strength of the electromechanical coupling is bounded above,
by the requirement that the energy changes involved in electron
tunnelling are dominated by the SET bias voltages rather then the
position dependent correction, i.e.\ $eV_{1(2)} \gg
 m\omega_0^2x_{1(2)}^2$, or equivalently $\kappa/\Gamma_{1(2)} \ll 1$. The
range of values of bias voltages is bounded above by the
requirement that the number of accessible charge states for each
SET should be limited to two, which essentially means that the
charging energy of the SET islands should be the largest
energy-scales in the problem, i.e.\ $e^2/2C_{1(2)\Sigma}\gtrsim
eV_{1(2)}$. In what follows, when we discuss variations in the
system parameters, we will implicitly be assuming that the
variation is always within the range discussed. For example, when
we consider the case where $V_1 \gg V_2$, we nevertheless assume
that $e^2/2C_{1(2)\Sigma}\gtrsim eV_1 \gg eV_2 \gg
m\omega_0^2x_1^2$.

\section{Resonator Dynamics} \label{sec:meq}

The dynamics of the resonator is governed by the evolution of the
probability distribution $P(x,u;t)=\sum_{n1,n2}
P^{n_1,n_2}(x,u;t)$. The equation of motion for $P(x,u;t)$ forms
part of an apparently complex set of four coupled equations [Eqs.\
(\ref{eq:motionP})-(\ref{eq:PN12})]. However, equations of motion
for moments of the resonator's probability distribution are
readily obtained from the master equations. Furthermore, the
master equations can be solved numerically using the techniques
described in Ref.\ [\onlinecite{ABZ}].

The behavior of the first moments of the resonator are described
by a set of four coupled differential equations:
\begin{eqnarray}
\dot{\langle x \rangle} &=& \langle u \label{eq:firstmom_t1}\rangle\\
\dot{\langle u \rangle} &=& - \epsilon^2 \left[\langle x \rangle -
\langle P \rangle_{N_1+1}+ \langle P
\rangle_{N_2+1}\right]\label{eq:firstmom_t2}\\
\dot{\langle P \rangle}_{N_1+1} &=& \frac{\Gamma_1}{2} + \kappa
\langle x \rangle - \Gamma_1\langle P
\rangle_{N_1+1}\label{eq:firstmom_t3}
\\
\dot{\langle P \rangle}_{N_2+1} &=&\frac{\Gamma_2}{2} - \kappa
\langle x \rangle - \Gamma_2 \langle P
\rangle_{N_2+1}\label{eq:firstmom_t4},
\end{eqnarray}
where the electron-resonator moments are defined using the
notation
\begin{eqnarray}
\langle \dots \rangle_{N_{1(2)}+1}&=&\sum_{n_1,n_2} \int dx \int
du
(\dots)P^{n_1,n_2}_{N_{1(2)}+1}(x,u;\tau) \nd \\
\av{P}_{N_{1(2)}+1}&=&\int\int dx du \sum\limits_{n_1,n_2}
{P^{n_1,n_2}}_{N_{1(2)}+1}.
\end{eqnarray}
The full solution of this set of equations offers little insight,
but if we make some approximations based on the realistic
assumption that the resonator frequency is much lower than the
electronic tunnelling rate (i.e.\ $\epsilon \ll 1$), then  a
simple equation of motion for the average position of the
resonator about its fixed point can be derived:
\begin{equation}
 \ddot{\av{x}} = -\epsilon^2\left(1-\frac{\kappa}{\Gamma_1}
  -\frac{\kappa}{\Gamma_2}\right)\av{x} -
 \kappa\epsilon^2\left( \frac{1}{{\Gamma_1}^2}+ \frac{1}{{\Gamma_2}^2}\right)
\dot{\av{x}}, \label{eq:damped}
\end{equation}
as discussed in the Appendix. It is clear that the resulting
equation of motion for the average position of the resonator is
essentially just that of a damped harmonic oscillator with a
shifted frequency,
$\epsilon'=\epsilon[1-\kappa(\Gamma_1^{-1}+\Gamma_2^{-1})]^{1/2}$,
and a damping constant $\gamma_{SET}=\gamma_1+\gamma_2$, where
$\gamma_{1(2)}=\kappa\epsilon^2/\Gamma^2_{1(2)}$.

The form of the equation of motion for $\av{x}$ strongly suggests
that the resonator will have a well-defined steady state, and
indeed this is what is found when the master equations [Eqs.\
(\ref{eq:motionP})-(\ref{eq:PN12})] are integrated numerically.
Furthermore, numerical integration also reveals that for
$\kappa/\Gamma_1,\kappa/\Gamma_2\ll 1$ the steady state
probability distribution is very well approximated by a Gaussian,
something that is readily verified by calculating the steady state
values for higher moments of the resonator. Thus we can
characterize the steady state of the resonator by the variances in
the position and velocity, $\delta x^2$ and $\delta u^2$.

We can calculate equations of motion for the second moments in a
similar way to that employed in Ref.\ \onlinecite{ABZ}. For
example the equation of motion for $\langle x^2\rangle$ is
obtained by by multiplying Eq.\ (\ref{eq:motionP}) by $x^2$ and
then integrating over $x$ and $u$ in the the same way as for the
first moments. Setting the rate of change of the second order
moments to zero (i.e.\ $\dot{\av{x^2}}=0$ etc.) leads to a coupled
set of algebraic equations for the steady state values of the
second moments. For $\epsilon \ll 1$, we find that the variances
$\delta x^2= \av{x^2}-\av{x}^2$ and $\delta u^2=
\av{u^2}-\av{u}^2$ obey equipartition. When we are also in the
weak coupling limit ($\kappa/\Gamma_{1(2)}\ll 1$) the variances
(in dimensionful units) are
\begin{equation}
m\delta u^2 =m\omega_0'^2\delta
x^2=\frac{e}{4}\frac{\Gamma_2^2V_1+\Gamma_1^2V_2}{\Gamma_1^2+\Gamma_2^2},
\end{equation}
where $\omega_0'=\epsilon'\Gamma$. Bearing in mind that if only
SET 1(2) were present the system would (in this weak coupling
limit) have an effective temperature of the form\cite{ABZ}
$eV_{1(2)}/4$, we can rewrite the variances in terms of an
effective temperature which is an average of the effective
temperatures associated with each of the SETs:
\begin{equation}
m\delta u^2=k_{\rm B}T_{SET}=k_{\rm B}\frac{T_1 \gamma_1+T_2
\gamma_2}{(\gamma_1+ \gamma_2)},
\end{equation}
where $k_BT_{1(2)}=eV_{1(2)}/4$. These results show that in the
weak-coupling limit the effective temperature and damping constant
for the resonator coupled to two SETs take exactly the same form
as those obtained previously for a resonator coupled to one SET
and an external heat bath, characterized by its own temperature
and damping constant.\cite{ABZ} Thus for the two-SET resonator
system we can think of each of the two SETs as acting on the
resonator as an independent heat bath.

For the particular case when the two SETs are identical (i.e.\
 $V_1=V_2$), then the ratio of the variances
of the resonator position when coupled to just one of the SETs,
$\delta x^2_{(1)}$, to that when it is coupled to both SETs,
$\delta x^2_{(2)}$, takes the form
\begin{equation}
\frac{\delta x^2_{(2)}}{\delta
x^2_{(1)}}=\frac{1-\kappa}{1-2\kappa}.
\end{equation}
For $\kappa\ll 1$ it seems that adding a second SET (identical to
the one already coupled to the resonator) has almost no effect on
the resonator's state. However, although the variances of the
resonator are almost unaffected by adding (or removing) an
identical SET, the value of the damping due to the SET electrons
doubles when a second SET is added.

In practice the dynamics of a nanomechanical resonator would also
be affected by thermal fluctuations arising from its surroundings
which can be characterized by a temperature, $T_{e}$ and damping
constant, $\gamma_e$. However, in what follows we will assume that
the dynamics of the resonator is dominated by the SET back-action
so that $\gamma_e\ll \gamma_{SET}$ and $T_e\ll T_{SET}$ and hence
neglect the effects of these additional fluctuations on the SET
noise characteristics. This back-action dominated regime is
expected to be accessible experimentally for resonators with
lengths $\sim 10\mu$m or longer, as discussed in Ref.\
\onlinecite{ABZ}.

\section{Noise Properties} \label{sec:current}

Motion of the resonator affects the tunnel rates of the electrons
in both the SETs and hence can induce correlations not just in the
current flowing in the individual SETs, but also between the
currents in the two SETs, known as auto- and cross-correlations
respectively. The power spectrum of the auto-correlation function
gives the current noise in one of the SETs, while the current
noise of the combined currents of the two SETs are given by a
combination of the power spectra of the auto- and
cross-correlations. We begin this section by outlining how the
zero frequency  current noise in one of the SETs of the two-SET
resonator system can be obtained. Then we analyze the noise in one
of two SETs coupled to a resonator, and compare it with the noise
in a single SET coupled to a resonator which is held in a fixed
thermal state rather than one determined by the back-action of the
SETs. Finally we examine the cross-correlations in the currents of
the two SETs induced by being coupled to a resonator, and hence
calculate the noise in their combined current.

\subsection{Current noise in one SET}
\label{sec:auto}

The zero frequency current noise of SET 1, $S_{I_1I_1}(0)$, is
equal to the noise in the tunnel current across either of its
junctions. Considering the source junction for concreteness, we
have\cite{gurvitz}
\begin{equation} \label{eq:SIIdef}
S_{I_1I_1}(0)=  2 e^2 \frac{d}{d \tau} \left[ \sum\limits_{n_1}
{n_1}^2 P^{n_1}(\tau) - \left( \sum_{n_1} n_1 P^n_1(\tau)
\right)^2 \right]_{\tau\rightarrow \infty},
\end{equation}
where
\[
P^{n_1}(\tau)=\sum_{n_2}\int dx\int duP^{n_1 n_2}(x,u;\tau).
\]
and we have used a result due to MacDonald\cite{mac} to rewrite
the current through the source junction, $I_s(t)$, in terms of the
number of charges that have passed through that
junction.\cite{gurvitz} Substituting for the rates of change of
the distributions from the master equations  [Eqs.
(\ref{eq:motionP})--(\ref{eq:PN12})], we can rewrite the
expression for the noise, in units of time where $\Gamma=1$, as:
\begin{equation} \label{eq:SIIdef2}
S_{I_1I_1}(0) = \lim_{\tau \rightarrow \infty}2 \left[ 2 \left\{
e^2 \left( \frac{\Gamma_{1}}{2} \avpo{n_1} -\kappa
\frac{\avpo{xn_1}}{x_1}\right) - I_1^2\tau\right\} +e
 I_1 \right].
\end{equation}

\begin{figure}
\noindent
\epsfig{figure=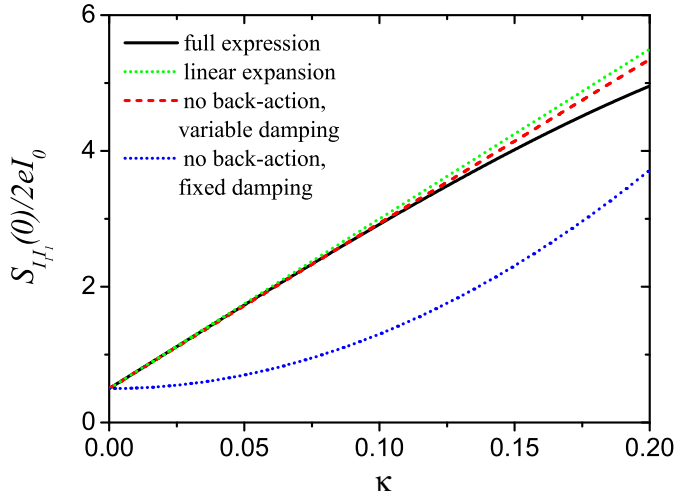,width=10cm}
 \caption{(Color online) Current noise through one of two SETs coupled to a
 resonator plotted as a function of the coupling strength, $\kappa$.
 The approximate expression for the current noise [Eq.\ \ref{eq:noise1}] and the noise of
a SET coupled to a resonator whose state is controlled solely by a
coupling to an external bath with fixed or variable damping are
also shown for comparison. For each plot we have set
$\epsilon=0.1$, and $V_1=V_2$} \label{fig:two_auto_exp}
\end{figure}

The calculation of the zero frequency noise thus reduces to the
calculation of the limiting behavior of two electron-resonator
moments. The long time limits of the moments are readily obtained
analytically using an extension of the equation of motion method
discussed in Ref.\ \onlinecite{ADA}. Substituting these values for
the moments into Eq.\ (\ref{eq:SIIdef2}), gives an expression for
the zero-frequency current noise. However, the resulting
expression is hard to interpret. Hence in order to obtain a more
accessible expression, we expand in the coupling parameter
$\kappa=m\omega_0^2x_0^2/eV$ and then make the simplifying
assumption $\epsilon\ll 1$. Up to order $\kappa$, the resulting
expansion is:
\begin{equation} \label{eq:noise1}
\frac{S_{I_1I_1}(0)}{2eI_0}=\frac{1}{2} +
\frac{{\Gamma_1}^2{\Gamma_2}^4(\Gamma_1+\Gamma_2)^2}
{2({\Gamma_1}^2+\Gamma_2^2)^3}\frac{\kappa}{\epsilon^2},
\end{equation}
where we have scaled the noise by $I_0$, the average current of
the SET without coupling to the resonator. The first term in the
expansion, $1/2$, is just the Fano factor of SET 1 (which in this
case has equal tunnel rates across both junctions) in the
non-interacting limit. The expansion is compared with the full
analytical expression in Fig.\ \ref{fig:two_auto_exp}; the
agreement is very good for $\kappa\ll 1$ as one would expect.

The presence of the resonator affects the current noise because
fluctuations in its motion, caused by interactions with the
electrons in either SET, generate correlations in the current. The
expansion in terms of $\kappa$ shows clearly that the current
noise increases with increasing coupling, measured by $\kappa$.
However, the expansion also shows that the noise is extremely
sensitive to the value of $\epsilon$, diverging in the limit
$\epsilon\rightarrow 0$. The explanation for this behavior is that
the zero frequency current noise is controlled by the fluctuations
in the state of the resonator which decay on a time-scale of order
the damping time, $1/\gamma_{SET}$. This damping time is
determined by the back-action, and also diverges in the limit
$\epsilon \to 0$. The fact that the intrinsic damping of the
resonator, $\gamma_{SET}$, vanishes as $\epsilon \to 0$ means the
fluctuations do not decay and therefore the current noise
diverges. This idea can be tested directly by comparing the
current noise of the two-SET resonator system with that of a SET
coupled to a resonator whose damping (and temperature) is
determined solely by coupling to an external thermal bath (i.e.
the back-action of the SETs on the resonator is neglected).

The back-action of the SET on the resonator arises from the
displacement of the equilibrium position of the resonator caused
by changes in the occupancy of the SET islands. When these
displacements are neglected, the motion of the resonator can still
affect the electron tunnelling rates of the SET, but the resonator
is no longer affected by the electrons. We calculate the zero
frequency noise in the same way as before but without the
back-action terms, instead assuming that the resonator is coupled
to an external equilibrium thermal bath, characterized by a
temperature $T_e$ and a damping constant, $\gamma_e$. With all
other parameters held constant, the current noise diverges as
$1/\gamma_e$, showing that the divergence in the current noise is
indeed due to the damping going to zero.\cite{blanter}

The noise obtained for the fully coupled system and the no
back-action model are compared in Fig.\ \ref{fig:two_auto_exp}.
Two particular choices of parameters for the external bath are
shown: one with a fixed value of the bath damping, $\gamma_e$, and
one with $\gamma_e$ chosen to match the $\kappa$ dependent
effective value of the equivalent fully coupled case (i.e.\
$\gamma_e=\gamma_{SET}$). In both cases, we choose the temperature
of the bath to match the small-coupling value of the equivalent
fully coupled case (i.e.\ we set $T_e=T_{SET}$, which is
independent of $\kappa$). The noise for the no back-action case is
almost linear in $\kappa$ when the value of $\gamma_e$ is varied
to match that of the equivalent fully coupled case, but only has a
small linear term when instead $\gamma_e$ is kept constant. This
shows that in the fully coupled case, it is the fact that the
damping depends on $\kappa$ that gives a linear (rather than
$\kappa^2$) dependence of the current noise.

\subsection{Current cross-correlations} \label{sec:cross}

Having analyzed the zero-frequency current noise of one of the
SETs in the two-SET resonator system, we now turn to consider the
correlations between the currents in the two SETs induced by the
presence of the resonator. The cross-correlation between  the two
SETs is defined as
\begin{equation}
K_{I_{1}I_{2}}(\tau)= \frac{1}{2}\left[
\av{{\tilde{I}}_1(t){\tilde{I}}_2(t+\tau)}+\av{{\tilde{I}}_2(t){\tilde{I}}_1(t+\tau)}
\right]
\end{equation}
where $\tilde{I}_i(t)=I_i(t)-\av{I_i}$, and the correlation
function is assumed to be independent of $t$. The power spectrum
of the cross-correlation is defined as,
\begin{equation}
S_{I_{1}I_{2}}(\omega)=2\int\limits_{-\infty}^\infty
\cos(\omega\tau)K_{I_{1}I_{2}}(\tau) \, d \tau.
\end{equation}
The zero frequency limit is independent of whether we consider
cross-correlations in the full currents flowing through the SETs
or in the tunnel currents across either the source or drain
junctions, and is given by
\begin{eqnarray}
S_{I_{1}I_{2}}(0)&=&S_{I_{1D}I_{2D}}(0)=S_{I_{1S}I_{2S}}(0)\\
&=&2e^2\frac{d}{d\tau}\left[\sum_{n_1}\sum_{n_2}n_1n_2P^{n_1n_2}(\tau)
-\left(\sum_{n_1}n_1P^{n_1}(\tau)\right)\left(\sum_{n_2}n_2P^{n_2}(\tau)\right)\right]_{\tau\rightarrow
\infty},
\end{eqnarray}
where we have used a generalization of MacDonald's formula in the
last line to recast the expression into a form which is readily
evaluated using the generalized master equations. Substituting
appropriately from the master equations [Eqs.\
(\ref{eq:motionP})-(\ref{eq:PN12})] for the rates of change of the
probability distributions, we obtain
\begin{equation} \label{eq:zerofreqnoisecross}
S_{I_{1}I_{2}}(0)=\lim_{\tau \rightarrow \infty} 2 \bigg\{
e^2\left( \frac{\Gamma_1}{2}\avpo{n_2} -
\kappa\avpo{xn_2}\right)\nonumber  +e^2\left(
\frac{\Gamma_2}{2}\avpt{n_1} -
\kappa\avpt{xn_1}\right)-2I_{1}I_{2}\tau \bigg\}.
\end{equation}
The moments $\avpt{n_1}$ ($\avpo{n_2}$) and $\avpt{xn_1}$
($\avpo{xn_2}$) describe the correlations between the resonator,
the charge on SET 2 (1) and the charge that has passed through SET
1 (2). They are readily obtained using the same equation of motion
approach employed for the single-SET moments.

\begin{figure}
\noindent
\epsfig{figure=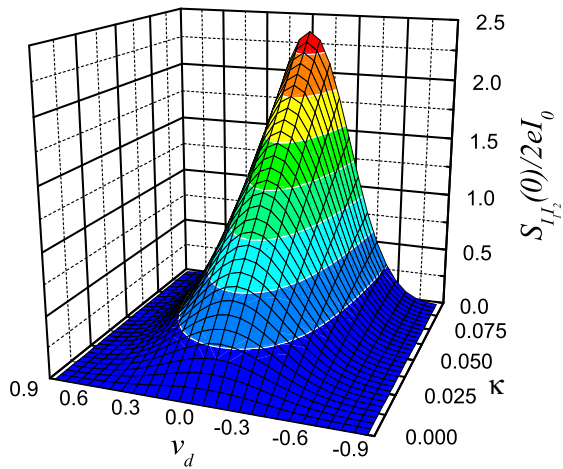,width=10cm} \caption{(Color online) The
zero frequency component of the cross-correlation power spectrum,
plotted as a function of $\kappa$ and $v_d$, with $\epsilon=0.1$.}
\label{fig:crossnoise2}
\end{figure}

The cross-correlation noise, $S_{I_1I_2}(0)$ is shown in Fig.\
\ref{fig:crossnoise2} as a function of the parameters $\kappa$ and
$v_d = (V_1-V_2)/(2V)$, which measures the difference between the
voltages. It is clear from the plot that the cross-correlations
are positive, reaching a maximum when the two voltages are equal
and increasing with the coupling, $\kappa$.

\begin{figure}
\noindent
\epsfig{figure=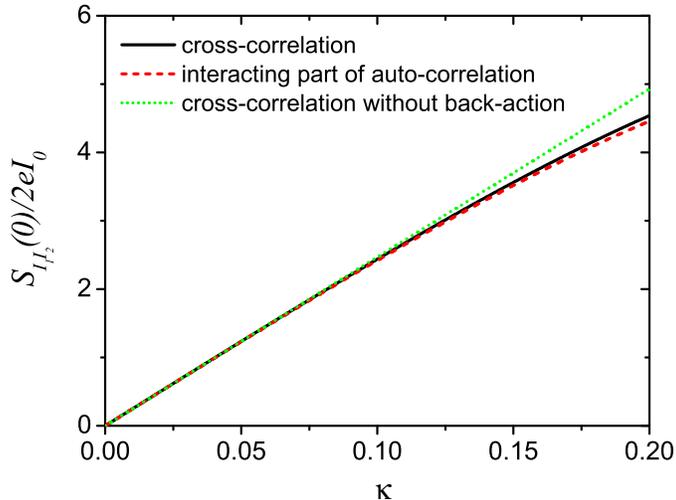,width=10cm} \caption{(Color online) A
comparison of the cross-correlation between the two leads, and the
auto-correlation minus the constant term
($S^-_{I_1I_1}(0)/(2eI_0)$) plotted as a function of $\kappa$ with
$\epsilon=0.1$ and $v_d=0$. Also plotted is the cross-correlation
without back-action.} \label{fig:cross_vs_auto}
\end{figure}

In order to understand the cross-correlation, we can again make a
comparison with the case of a resonator in a fixed thermal state
where the back-action of the electrons on the resonator is
neglected. Without the back-action, with the state of the
resonator controlled by coupling to an external equilibrium
thermal bath, there is no way for the electrons in the two SETs to
influence each other as they cannot affect the state of the
resonator. The zero frequency component of the cross correlation
with and without the back-action are compared in Fig.\
\ref{fig:cross_vs_auto}, where the temperature and damping
associated with the external bath for the no back-action case have
been chosen to match those of the fully coupled system, and it is
clear that they coincide up to linear order in $\kappa$. We can
think of the electrons in both SETs acting on the resonator to
form a dynamical state which then acts back on the electrons in
both SETs to induce correlations. When the back-action of the
electrons on the resonator is neglected, and it is kept in a
thermal state by an external bath, correlations are still induced
between the SETs because both SETs experience the {\it same}
fluctuations in the resonator's state. Fig.\
\ref{fig:cross_vs_auto} also compares the zero frequency component
of the cross-correlation with the corresponding part of the
auto-correlation which arises from interactions with the
resonator, $S^-_{I_1I_1}(0)/(2eI_0)=S_{I_1I_1}(0)/(2eI_0)-1/2$
[See Eq.\ (\ref{eq:noise1})], for identical SETs (i.e.\ $v_d=0$).
It is clear from  Fig.\ \ref{fig:cross_vs_auto} that
$S_{I_1I_2}(0)$ and $S^-_{I_1I_1}(0)$ are closely related even for
values of $\epsilon \sim 0.1$ which approach the limit of the
model's validity and indeed we find that in the limit $\epsilon
\rightarrow 0$ they become equal.

These results are at first a little surprising as the
cross-correlations between carriers in different leads of a device
are very often negative. Indeed, so long as the voltages applied
are constant and there are no interactions the cross-correlations
must be negative.\cite{buttiker_92,bandb,cottet_04} However, in
our system although the leads are kept at fixed voltages and
contain non-interacting electrons, the presence of the resonator
generates effective interactions within the device itself (i.e.\
between the islands of the SETs) and hence positive correlations
between currents in different leads are possible. Positive
current-correlations between leads were also found for similar
systems consisting of parallel quantum dots with a direct
electrostatic interaction between them.\cite{set1,note}

We might also expect the cross-correlations to be negative because
the signs of the SET-resonator coupling terms are opposite
$x_1=-x_2$ so that when the resonator moves towards SET 1
(effectively decreasing the tunnelling rate $\Gamma_{1S}^+$) it
moves away from SET 2 (increasing $\Gamma_{2S}^+$). However, we
must remember that the correlations are between ${I}_{1S}(t)$ and
${I}_{1S}(t+\tau)$ \emph{in the limit $\tau \rightarrow \infty$},
i.e.\ all motions of the resonator are averaged out over a long
time. The two SETs experience the same fluctuations in the state
of the resonator and hence, provided the SETs are identical, these
fluctuations induce essentially the same correlations between the
currents of the two SETs as within the current of each one
separately.
 However, this will not
be the case at finite frequency. In particular, we would expect
strong anti-correlation between the currents in the two SETs at
the resonator frequency.

A simple way to measure the cross-correlations in the currents
through the two SETs is to measure the noise in the combined
current of the two SETs, $I_T=I_1+I_2$, and compare the result
with the noise in the currents through the individual SETs, $I_1$
and $I_2$. The zero frequency noise of the combined current is
given by
\begin{equation}
S_{I_TI_T}(0)=\left[S_{I_1I_1}(0)+S_{I_2I_2}(0)+2S_{I_1I_2}(0)\right].
\end{equation}
Thus measuring the noise in the combined current of the SETs and
comparing it with measurements of the noise in the individual SETs
provides a straightforward way to obtain the magnitude of the
resonator-induced cross-correlations.


\section{Conclusions} \label{sec:conclusion}

We have used a classical master equation approach to investigate
the noise properties of two SETs coupled to a nanomechanical
resonator in the regime where the dynamics of the resonator is
dominated by the back-action of the SETs. Although a classical
treatment of the resonator clearly cannot describe the ultimate
limits set by quantum mechanics on displacement
detection,\cite{mset1,mset2,white,Clerk2} it does give important
insights into the measurement back-action.

We have shown that provided the electromechanical coupling is
sufficiently weak ($\kappa \ll 1$) and the resonator frequency is
sufficiently low ($\epsilon\ll 1$), the SETs act on the resonator
like two independent thermal baths, each of which can be
characterized by a damping constant and effective temperature,
leading to an overall effective temperature for the resonator that
is an average of the effective temperatures associated with the
individual SETs.

The coupling to the resonator generates positive correlations
between the electrons flowing through each of the SETs
individually and hence leads to an enhancement of the low
frequency current which indeed can become parametrically large.
The noise depends sensitively on both the SET resonator couplings
and the effective damping rate of the resonator which arises from
interactions with the SET electrons. In particular, the zero
frequency noise diverges as the damping rate of the resonator
tends to zero. For sufficiently weak coupling, we found that the
magnitude of the noise matches that which would arise from
coupling to a resonator in a fixed thermal state characterized by
an appropriately chosen temperature and damping rate. These
results imply that the enhancement of the low frequency noise in
the SETs can be thought of as due to fluctuations in the dynamical
state of the resonator which persist for times of order the
resonator's damping time.

 The presence of
the resonator also generates positive correlations between the
currents in the two SETs. These correlations would be manifest in
the current noise of the combined current of the two SETs which
would be greater than the sum of the noise in the individual SET
currents. Furthermore, the low frequency limit of the spectra of
the cross- and auto-correlations induced in the SET currents by
the resonator are almost identical for $\epsilon \ll 1$.

\section*{Acknowledgements}
We thank M.P. Blencowe for a series of very useful discussions.
This work was supported by the EPSRC under grant GR/S42415/01.

\appendix

\section{Solution of mean-coordinate equations for resonator}

In this Appendix we derive the effective equation of motion for
the mean-coordinate of the resonator [Eq.\ (\ref{eq:damped})]. The
derivation relies on a separation of the time scales associated
with the electrical and mechanical motion such that $\epsilon\ll
1$. The electron distribution almost relaxes to its equilibrium
value for {\it each} position of the resonator. This electron
distribution then acts on the resonator, leading to a frequency
shift. This is closely analogous to the Born-Oppenheimer
approximation describing the motion of atomic nuclei and electron
wavefunctions\cite{BO} (see Beausoleil \emph{et. al}\cite{beau_04}
for a similar analogy). However, rather than assume the resonator
moves infinitely slowly compared to the electrons (i.e.
$\epsilon=0$), we allow the resonator a small but finite
velocity.\cite{butler_98} As a consequence, we find that the
evolution of the electron distributions $\avpo{P}$, $\avpt{P}$
depend on $\av{u}$ as well as $\av{x}$. As we shall see, it is
this that leads to the damping effect.

We begin be rewriting the coupled equations for the first moments
of the SET-resonator distributions [Eqs.\
(\ref{eq:firstmom_t1})-(\ref{eq:firstmom_t4})] in terms of
variables centered on the appropriate fixed point value, e.g.\
$\langle {\tilde{P}} \rangle_{N_1+1} =\avpo{P}-\avpo{P}^{fp}$, and
hence obtain
\begin{eqnarray}
\dot{\av{\tilde{x}}} &=& \av{u} \label{eq:firstmom_s1}\\
 \dot{\av{u}}  &=& -\epsilon^2 \left( \av{\tilde{x}} - \tpo\right)\label{eq:firstmom_s2} \\
\langle \dot{\tilde{P}} \rangle_{N_1+1} &=&
\kappa\av{\tilde{x}} - \Gamma_1 \tpo \label{eq:firstmom_s3}\\
\langle \dot{\tilde{P}} \rangle_{N_2+1} &=& -\kappa \av{\tilde{x}}
- \Gamma_2\tpt\label{eq:firstmom_s4}.
\end{eqnarray}
From Eq. \ref{eq:firstmom_s3}, we have:
\begin{eqnarray}
\langle \dot{\tilde{P}} \rangle_{N_1+1} &=&
 - \Gamma_1\tpo +\kappa \av{\tilde{x}}\\
\langle \ddot{\tilde{P}} \rangle_{N_1+1} &=&{\Gamma_1}^2\tpo
 - \Gamma_1 \kappa \av{\tilde{x}}+\kappa\av{u} \label{eq:approxpddot2}\\
\langle \dddot{\tilde{P}} \rangle_{N_1+1} &=& -{\Gamma_1}^3\tpo
 + {\Gamma_1}^2\kappa\av{\tilde{x}}-\Gamma_1\kappa\av{u}+\kappa \epsilon^2( \tpo - \tpt -\av{\tilde{x}})\nonumber \\
 &\simeq & -\Gamma_1\langle \ddot{\tilde{P}} \rangle_{N_1+1}.\label{eq:approxpddot}
\end{eqnarray}
Where the approximate expression on the last line arises when we
neglect terms of order $\epsilon^2$, which we are assuming to be
small.

Solving Eqs. (\ref{eq:approxpddot}) and (\ref{eq:approxpddot2})
for $\tpo$, we  obtain
\begin{equation}
\tpo (t) =
\frac{\av{\tilde{x}}}{\Gamma_1}\kappa-\frac{\av{u}}{{\Gamma_1}^2}\kappa
+e^{-\Gamma_1t}C,
\end{equation}
where $C$ is a constant that depends on the initial conditions. If
we consider a timescale long compared to the motion of the
electrons, but short compared to the motion of the resonator (i.e.
neglect the transient behavior), then the second term can be
dropped and we obtain expressions for $\tpo$  as a function of
$\av{x}$ and $\av{u}$:

\begin{equation}
\langle \tilde{P} \rangle_{N_{1}+1} (t\gg1/\Gamma_1) =
\frac{\av{\tilde{x}}}{\Gamma_{1}}\kappa-\frac{\av{u}}{\Gamma_{1}^2}\kappa
\label{eq:Pfpxu1}.
\end{equation}
Obtaining a similar expression for $\tpt$, and inserting these
into Eqs. (\ref{eq:firstmom_s1}) and (\ref{eq:firstmom_s2}) gives
the effective equation of motion for the resonator quoted in the
main text:
\[
 \ddot{\av{\tilde{x}}} = -\epsilon^2\left(1-\frac{\kappa}{\Gamma_1}-\frac{\kappa}
 {\Gamma_2} \right)\av{\tilde{x}} -
 \epsilon^2\left( \frac{\kappa}{{\Gamma_1}^2}+\frac{\kappa}{{\Gamma_2}^2} \right)
\dot{\av{\tilde{x}}}.
\]

Notice that we needed to allow $\tpo$ to depend on both $\av{x}$
and $\av{u}$ in order to obtain this effective equation of motion.
If instead, we had included only the $\av{x}$ dependence, we would
have obtained only the frequency shift and not the damping. The
sign of the $u$ term determines whether we have a damping term or
one that drives the oscillator further from
equilibrium.\cite{novotny} Including higher order derivatives in
the calculation of $\tpo$ leads to additional terms in the
expressions for the frequency shift and damping of order
$\epsilon^4$.


\end{document}